\documentclass[twocolumn]{aastex61}
\usepackage{graphicx,subfigure,amsmath, amsfonts, amssymb,footnote,color,epstopdf}
\usepackage{apjfonts}
\begin{document}

\title{ALMA observations of the interaction of a radio jet with molecular gas in Minkowski's Object}


%

\correspondingauthor{Mark Lacy}
\email{mlacy@nrao.edu}

\author{Mark Lacy}
\affiliation{National Radio Astronomy Observatory, Charlottesville, VA 22903, USA}

\author{Steve Croft}
\affiliation{Astronomy Department, University of California, Berkeley, 501 Campbell Hall \#3411, Berkeley, CA 94720, USA}
\affiliation{Eureka Scientific Inc, 2452 Delmer St Suite 100, Oakland, CA 94602}

\author{Chris Fragile}
\affiliation{Department of Physics and Astronomy, College of Charleston, 66 George Street, Charleston, SC 29424}

\author{Sarah Wood}
\affiliation{National Radio Astronomy Observatory, Charlottesville, VA 22903, USA}

\author{Kristina Nyland}
\affiliation{National Radio Astronomy Observatory, Charlottesville, VA 22903, USA}


\begin{abstract}
We use ALMA to detect and image CO (1-0) emission from Minkowski's Object, a dwarf galaxy that is interacting with a radio jet from a nearby elliptical galaxy. These observations are the first to detect molecular gas in Minkowski's Object. We estimate the range in the mass of molecular gas in Minkowski's Object assuming two different values of the ratio of the molecular gas mass to the CO luminosity, $\alpha_{\rm CO}$. For the Milky Way value of $\alpha_{\rm CO}=4.6~M_{\odot}{\rm (K~km~s^{-1}~pc^2)^{-1}}$ we obtain a molecular gas mass of $M_{\rm H_2} =3.0 \times 10^7~M_{\odot}$, 6\% of the H{\tt I} gas mass. We also use the prescription of Narayanan et al. (2012) to estimate an $\alpha_{\rm CO}=27~M_{\odot}{\rm (K~km~s^{-1}~pc^2)^{-1}}$, in which case we obtain $M_{\rm H_2} =1.8 \times 10^8~M_{\odot}$, 36\% of the H{\tt I} mass. The observations are consistent with previous claims of star formation being induced in Minkowski's Object via the passage of the radio jet, and it therefore being a rare local example of positive feedback from an AGN. In particular, we find highly efficient star formation, 
with gas depletion timescales $\sim 5\times 10^7 - 3\times 10^8$yr (for assumed values of $\alpha_{\rm CO}=4.6$ and $27~M_{\odot}{\rm (K~km~s^{-1}~pc^2)^{-1}}$, respectively) in the upstream regions of Minkowski's Object that were struck first by the jet, and less efficient star formation downstream. We discuss the implications of this observation for models of jet induced star formation and radio mode feedback in massive galaxies.
\end{abstract}




\section{Introduction} \label{sec:intro}

Active Galactic Nucleus (AGN) feedback by radio jets is usually thought of as strongly negative, reducing the ability of the host galaxy to form stars by heating or injecting turbulence into the interstellar medium of the host galaxy (e.g.\ Nesvadba et al. 2007, 2016; Alatalo et al. 2015) or the surrounding intracluster medium (e.g. Br\"{u}ggen \& Kaiser 2002; Ma et al. 2013). Negative feedback can also lead to expulsion of gas from the galaxy, though typically only a small fraction of gas exceeds the escape velocity (e.g.\ Dasyra et al. 2014). Nevertheless, there is a significant observational evidence of positive feedback through jet induced star formation, principally as an explanation of the ``alignment effect'' in high redshift, high luminosity radio-loud AGN, where the radio and optical structures frequently appear co-aligned (Chambers, Miley \& van Breugel 1990, Best, Longair \& R\"{o}ttgering 1996, Dey et al. 1997; Lacy et al. 1999; Clements et al. 2009; Zinn et al. 2013; Gullberg et al. 2016). Jet induced star formation has also been suggested in several lower redshift objects, e.g.\ Mould et al. (2000), Osterloo \& Morganti (2005), Crockett et al. (2012) and Santoro et al. (2016) in Centaurus A;  Elbaz et al. (2009) in a companion to the ``hostless'' $z=0.286$ quasar HE0450-2958, as well as in Minkowski's Object, the subject of this paper.

Theoretical studies, initially driven by the discovery of the alignment effect, have also proposed jet induced star formation as an important mechanism for star formation in AGN (Rees 1989; De Young 1989). Wagner et al. (2016) suggest that powerful radio jets can provide both positive feedback (via jet-induced star formation) and negative feedback in their hosts. Bieri et al. (2015) make the case that jet-induced star formation  could be responsible for the very high star formation rates seen in some high redshift galaxies. Gaibler et al. (2012), Dugan et al. (2016) and Fragile et al. (2017) suggest that jet induced star formation may produce characteristic ring morphologies in the young stellar population. Most of these studies have concentrated on high power FRII radio sources, and the existence and importance of these processes in the lower luminosity radio galaxies is unclear. Fragile et al. (2004; 2017) modeled the jet-cloud interaction in Minkowski's Object, and were able to show that cooling in the post-shocked gas can be effective at forming molecular gas and stars. Mukherjee et al. (2016)  suggest that, because they spend longer confined in their host galaxies, lower power radio jets may have a greater effect on the ISM than high power ones. This is important to establish, as low power radio galaxies are key to the "radio" or "maintenance" mode of negative feedback that is thought to be the dominant mechanism for the cutoff in star formation in the most massive galaxies (Croton et al. 2006). Silk \& Nusser (2010) make the case that, paradoxically, jet induced star formation may actually enhance negative feedback effects, as the supernovae produced by triggered star formation may enhance the momentum in the outflow of gas from the nucleus. This provides a mechanism by which the relatively low power radio jets can couple effectively to the ISM. Thus proper understanding of the nature and role of jet induced star formation may be critical to our understanding of both negative and positive feedback.  

Minkowski's Object is a star forming dwarf galaxy in the poor cluster Abell 194, and a rare low redshift example of a radio jet interacting with a galaxy that is not its host. This allows us to make a detailed, spatially resolved study of the interaction. Jet induced star formation has long been suggested as a possible mechanism for triggering the star formation (Brodie et al. 1985, van Breugel et al. 1985), but the evidence remains circumstantial. To date, the definitive study of this object is that of Croft et al. (2006; hereafter C06), who use radio continuum, HI, ultraviolet, optical and near-infrared data to make a convincing case for triggering of star formation by the radio jet in this object.

The origin of Minkowski's Object remains unclear. The cluster is the host of two FRI radio galaxies, one, hosted by NGC541 (3C40A), is the galaxy responsible for the jet impacting Minkowski's Object, but the other (3C40B), hosted by the nearby galaxy NGC547 (itself one of a close pair of galaxies with NGC545) is of slightly higher radio luminosity (Sakelliou, Hardcastle \& Jetha 2008). These interactions, and the linear shape of Abell 194 (Bliton et al. 1998), suggest that it is quite a young cluster, possibly in the process of virialization. There are thus two likely scenarios for the origin of Minkowski's Object, (1) it was a pre-existing dwarf galaxy in the cluster that happened to fall into the path of a radio jet, or (2) it formed in situ from a gas cloud between NGC545/547 and NGC541, perhaps placed there as a result of tidal forces during the interaction (deep optical images show a stellar bridge between NGC545/547 and NGC541, along which Minkowski's Object is found (C06)). If Minkowski's Object is a pre-existing dwarf, then, given the rarity of gas-rich dwarf galaxies near the centers of clusters is is probably on its first passage through the cluster (e.g.\ Tran et al. 2005).


The interstellar medium (ISM) of Minkowski's Object was well studied in the atomic and ionized phases in C06, however, the molecular gas that is the fuel for star formation has remained undetected prior to this study. The low metallicity and low molecular gas mass of Minkowski's Object has made it a challenge to detect in CO (though an attempt was made with the Institut de Radioastronomie Millim\'{e}trique (IRAM) 30m telescope by Salom\'{e} et al. (2015)). We therefore requested time on the Atacama Large Millimeter Array (ALMA) to obtain a detection of the CO emission from Minkowski's Object, and obtained data in Cycle 3 (program 2015.1.00570.S).  In this paper we describe the results of these observations, the first observational study in a series of new observations and numerical simulations (Fragile et al.\ 2017) of Minkowski's Object that we expect to significantly enhance our understanding of the role of positive feedback in galaxy formation.

We assume a cosmology with $H_0=70 {\rm kms^{-1} Mpc^{-1}}$, $\Omega_{\Lambda}=0.7$ and  $\Omega_{\rm M}=0.3$ and a redshift of $z=0.0189$ for Minkowski's Object, at 
which the luminosity distance is 82Mpc and 1-arcsec corresponds to 381pc. All velocities are quoted in the ``radio'' convention, $\frac{v}{c}=\frac{(\nu_{\rm rest}-\nu_{\rm obs})}{\nu_{\rm rest}}$.

\section{Data}

\subsection{ALMA Observations}
Minkowski's Object was observed over four executions of a single scheduling block on 2016 January 2 and 2016 January 3.  The total time on source was 173 minutes.  Four spectral windows were observed with a bandwidth of 1875~MHz each, one centered on the expected velocity of the CO(1-0) emission line at 113.15~GHz, and three (centered at 99.2, 101.1 and 111.2~GHz) to provide continuum.  Observatory calibration was used, with Uranus being observed as the flux density calibrator, this should result in flux densities accurate to $\approx 5$\%.  The spectral resolution is 3.0~km~s$^{-1}$, and the 50\% power point of the primary beam is at a radius of $28^{\prime \prime}$. The data were calibrated using the ALMA pipeline and imaged using the Common Astronomy Software Applications (CASA) package (McMullin et al. 2008).  We formed the continuum image using the multifrequency synthesis mode of the {\sc clean} task, employing two Taylor terms.  In order to maximize the sensitivity to low surface brightness emission we used natural weighting during imaging.  This resulted in an image with a mean effective frequency of 106~GHz, a synthesized beam size of $3.5^{\prime \prime} \times 2.6^{\prime \prime}$ at a position angle of 275$^{\circ}$, and an rms noise of 60~$\mu$Jy~beam$^{-1}$. A primary beam correction was applied to the continuum image when making the spectral index map (Section 3.2), but was not applied to the intensity images in Figures 2 and 3, where the 50\% power point of the primary beam is indicated by the black dashed circles. To obtain a CO cube, we used the CASA task {\sc uvcontsub} to model and subtract the continuum emission using the line-free channels. The synthesized beam size in the CO(1-0) image is $3.2^{\prime \prime} \times 2.4^{\prime \prime}$ at a position angle of 275$^{\circ}$ and the rms noise in a single 3kms~$^{-1}$ channel is 0.5 mJy~beam$^{-1}$. As the CO emission is completely contained within the 95\% power point of the primary beam, no primary beam correction was applied to the CO images. We successfully detected, for the first time, CO emission from molecular gas, as well as continuum emission at 106~GHz associated with the radio jet.    

\subsection{Archival VLA data}
We retrieved VLA continuum data (project AV0102) observed in the B and C configurations with central observing frequencies of 1.44 and 1.63~GHz.  Standard calibration was performed in CASA.  We combined and imaged these data using multifrequency synthesis with two Taylor terms, resulting in an image at a mean effective frequency of 1.5~GHz. The final image has an rms noise of 0.15 mJy~beam$^{-1}$ and a synthesized beam size of $4.2^{\prime \prime} \times 3.8^{\prime \prime}$ at a position angle of 14$^{\circ}$.

C-configuration H{\tt I} observations of Minkowski's Object from program AM789 (observed 2004 April 2 for $\approx 4.3$hr time on source) were presented in C06. We extracted these data from the archive, along with unpublished B-configuration from program AB1159 (observed 2005 May 22-24 for a total of $\approx 15.6$hr on source), and calibrated and imaged the combined dataset in CASA. The calibrator for both observations was 3C48. One iteration of self calibration in phase only was applied. Continuum subtraction was performed in the image plane. To achieve sufficient sensitivity to the low surface brightness H{\tt I} emission we applied a $uv$-plane taper that resulted in a beam of $15.6^{\prime \prime} \times 15.0^{\prime \prime}$ at a position angle of -13$^{\circ}$. The final cube was imaged with a channel width of 6kms~$^{-1}$, at which the noise per channel was 0.5mJy~beam$^{-1}$.

\section{Results}

\begin{figure*}
\plotone{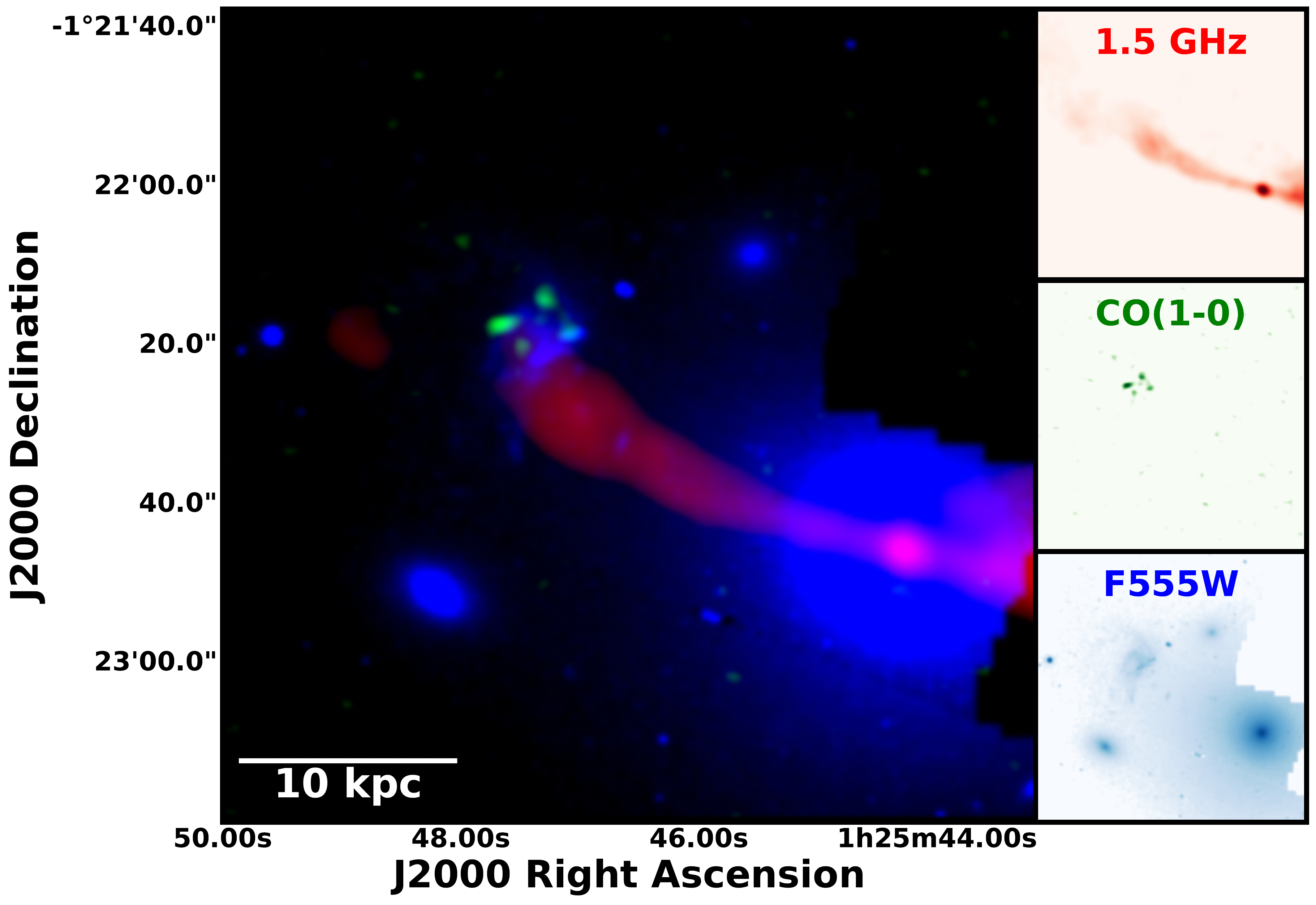}
\caption{Overview of the VLA, HST, and ALMA imaging of Minkowski's Object.  The 1.5~GHz VLA image of the radio jet is shown in red, the HST F555W image is shown in blue, and the CO(1-0) moment 0 map from our new ALMA spectral line observations is shown in green.  The large image is a three-color, rgb composite image, while the smaller images on the right highlight each band individually.
\label{fig:overview}}
\end{figure*}

\begin{figure*}
\centering
\includegraphics[clip=true, trim=0.2cm 0cm 0cm 0cm, width=8.95cm]{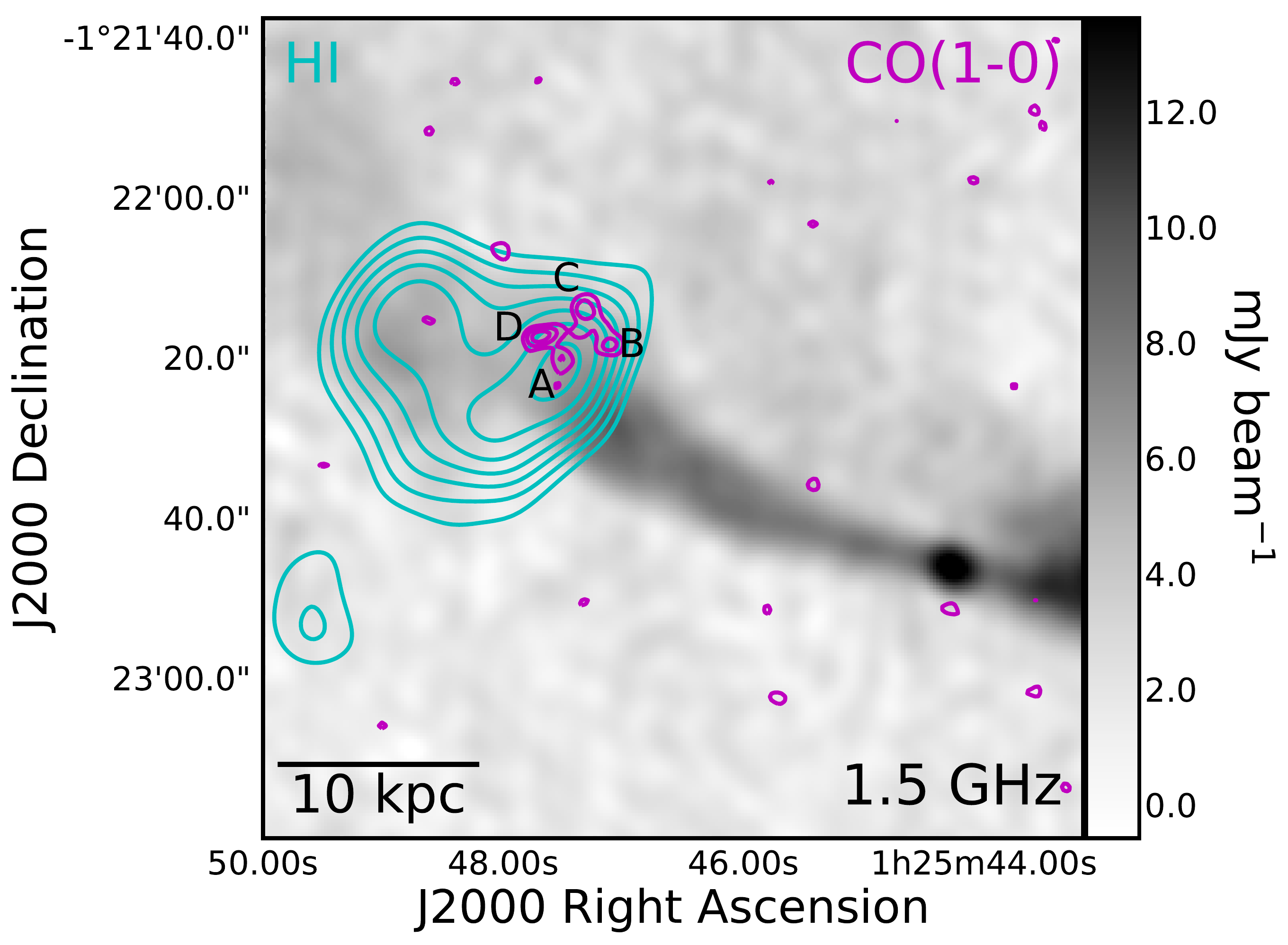}
\includegraphics[clip=true, trim=0cm 0cm 0.2cm 0cm, width=8.95cm]{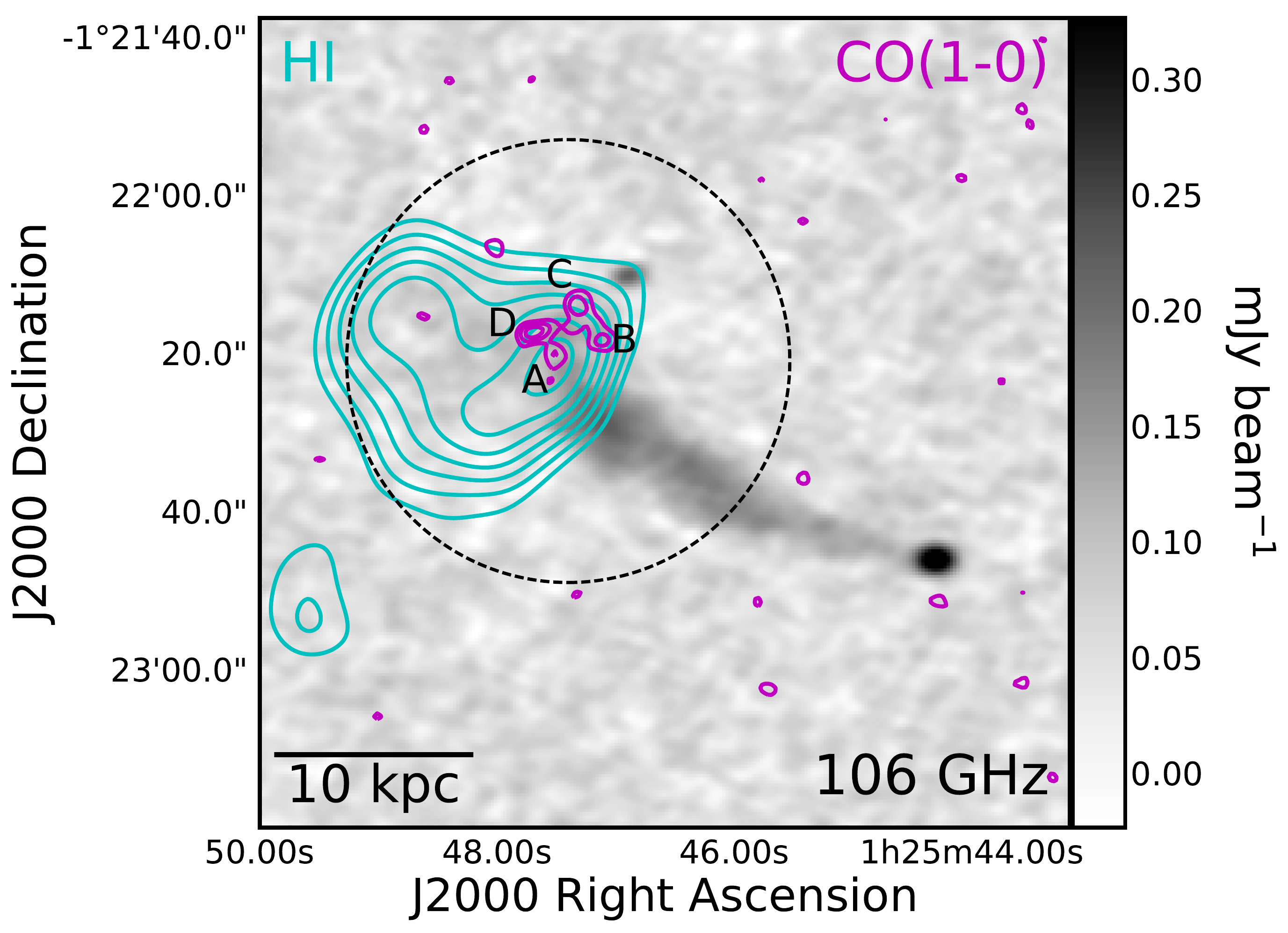}
\caption{ {\bf Left:} Overlay of the H{\tt I} contours from C06 (green) and the ALMA CO(1-0) contours from this work (cyan) on a grayscale image of the 1.5~GHz VLA continuum emission.  Both H{\tt I} and CO contours are drawn at intervals of 10~mJy~km~s$^{-1}$~beam$^{-1}$, starting at 20~mJy~km~s$^{-1}$~beam$^{-1}$ in CO and 50~mJy~km~s$^{-1}$~beam$^{-1}$ in H{\tt I}.  CO clumps A-D are labeled.  {\bf Right:} Same as the left panel, except here the grayscale image shows the ALMA 106~GHz continuum image, and the dashed black circle indicates the 50\% power point of the ALMA primary beam.} 
\label{fig:hi}
\end{figure*}

\subsection{CO emission}
In Figure~\ref{fig:overview}, we provide an rgb image of Minkoswki's Object.  This image shows our new ALMA detection of CO(1-0) emission in relation to the VLA centimeter- and ALMA millimeter-wave emission associated with the radio jet, as well as optical emission from HST imaging.
In Figure~\ref{fig:hi}, we show the CO(1-0) and H{\tt I} moment zero maps superimposed on the radio continuum images at 1.5 and 106~GHz. The CO emission appears clumpy with two clumps (A and B) seemingly closely associated with the star forming regions, a clump to the North (C), and a further one, D, to the East (in the downstream direction of the jet). The clumps are overlaid on a diffuse background of emission extending over most of the extent of the galaxy in the optical. 

\subsection{Continuum emission}
\begin{figure*}
\plotone{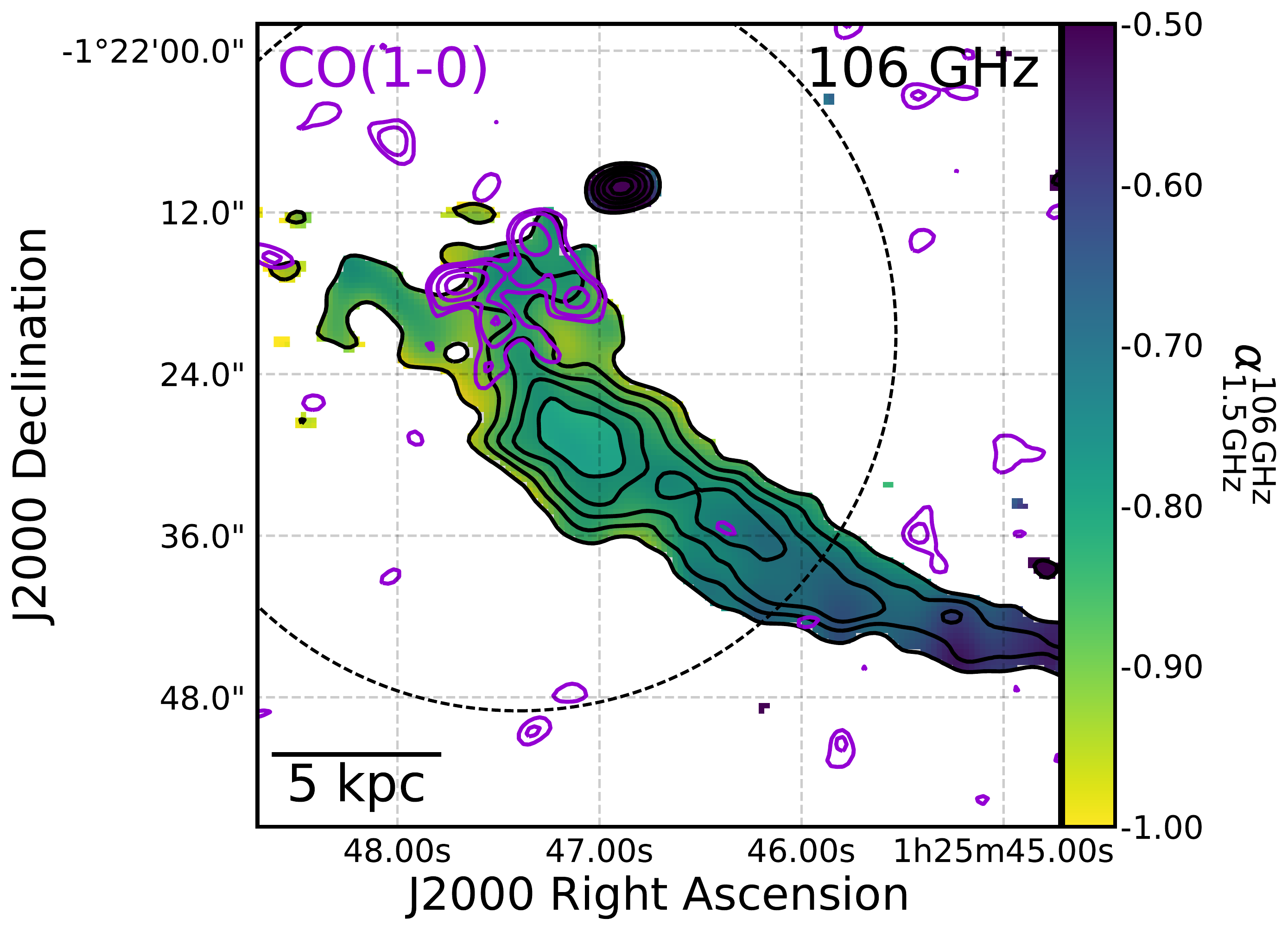}
\caption{Colorscale image of the spectral index between 1.5 and 106~GHz with contours of the ALMA 106~GHz continuum emission (black) and the ALMA CO(1-0) moment 0 image (purple). Contours for the 106~GHz image are drawn at 15~$\mu$Jy~beam$^{-1}$ intervals, starting at 15~$\mu$Jy~beam$^{-1}$, and for the moment 0 image at 15, 20, 30 and 40~mJy~km~s$^{-1}$~beam$^{-1}$.  CO clumps A-D are labeled. The dashed black circle indicates the 50\% power point of the ALMA primary beam.}
\label{fig:spix}
\end{figure*}

A radio spectral index\footnote{Here we follow the convention $S \propto \nu^{\alpha}$, where $S$ is the flux density, $\nu$ is the observing frequency, and $\alpha$ is the radio spectral index.} map between 1.5 and 106~GHz, made by convolving the ALMA image to the slightly lower resolution of the VLA image and then calculating the per-pixel spectral index, is shown in Figure \ref{fig:spix}.  This figure shows that the spectral index of the jet steepens from $\approx -0.6$ near the base of the jet to $\approx -0.8$ to $-0.9$ towards the edge of the jet where it is interacting with the molecular gas.  
Figure~\ref{fig:spix} also suggests a slight flattening of the spectral index to about $-0.77$ near the interaction with the easternmost CO clump D, consistent with a modest amount of re-acceleration of electrons via shocks in the interaction region (c.f.\ Giacintucci et al. 2008). 

The star formation rate in Minkowski's Object is much too low for the continuum emission at 106~GHz to arise from thermal processes. Based on the star formation rate of 0.47~M$_{\odot}$~yr$^{-1}$ (Salom\'{e} et al.\ 2015) and the scaling of Murphy et al. (2012) we expect a flux density of $\approx 80$~$\mu {\rm Jy}$ at 106~GHz, compared to a measured flux density of 500~$\mu$Jy within the 16$^{\prime\prime}$ integrated aperture defined in Table 1; similarly we do not expect a significant contribution from synchrotron emission due to the starburst, which should be much fainter than the thermal emission at 106GHz. Murphy et al. (2012) suggest $\approx$~76\% of the emission at 33GHz in nearby galaxies is thermal, and at 106GHz the fraction will be even higher because of the steep spectrum of the synchrotron emission.
In a future study we will use new VLA data to tighten the existing constraints on the spatially-resolved radio spectral index of Minkowski's Object using data in additional frequency bands with comparable angular resolution to the existing data presented here.

\begin{deluxetable*}{cccCCCC}
\tablecaption{Molecular Gas Properties of Minkowski's Object \label{tab:gas_props}}
\tablecolumns{7}
\tablenum{1}
\tablewidth{0pt}
\tablehead{
\colhead{Region} & \colhead{RA, DEC.} & \colhead{Aperture} & \colhead{$v_\mathrm{mean}$} & \colhead{$\Delta v_{\mathrm{FWHM}}$} & \colhead{$S_{\mathrm{CO(1-0)}}$} & \colhead{$M_{\mathrm{H2}}$} \\
\colhead{} & \colhead{(J2000)} & \colhead{(arcsec)} & \colhead{(km~s$^{-1}$)} & \colhead{(km~s$^{-1}$)} & \colhead{(mJy~km~s$^{-1}$)} & \colhead{(10$^7$M$_{\odot}$)} \\
\colhead{(1)} & \colhead{(2)} & \colhead{(3)} & \colhead{(4)} & \colhead{(5)} & \colhead{(6)} & \colhead{(7)}
}
\startdata
Clump A &01:25:47.50 -01:22:20.5& 3.5 & 5557$\pm$1         &  20$\pm$3    & 26$\pm$8      &  0.19-1.1  \\
Clump B& 01:25:47.10 -01:22:18.5&    3.5& 5553$\pm$1      &   14$\pm$3 &   30$\pm$8     &  0.22-1.3  \\
Clump C& 01:25:47.33 -01:22:14.0&    3.5&5574$\pm$1      &  16$\pm$3  &    33$\pm$8    & 0.25-1.4  \\
Clump D&01:25:47.7 -01:22:17.2 &  3.5 & 5557$\pm$1    &   11$\pm$2    &   38$\pm$8     &  0.28-1.7  \\ 
\hline
Integrated &01:25:47.40 -01:22:19.0 &16 &  5556$\pm$1        &   26$\pm$3    &  $405\pm 81$  & 3-18 \\
\enddata
\tablecomments{Column 1: CO(1-0) region.  Column 2: Position of the CO(1-0) region defined in Column 1.  Column 3: Aperture diameter used for summing the CO(1-0) flux.  Column 4: Mean velocity of the CO(1-0) emission.  Column 5: Full width at half maximum of the CO(1-0) emission.  Column 6: CO(1-0) integrated line intensity, $S_{\mathrm{CO(1-0)}}$.  For the value of $S_{\mathrm{CO(1-0)}}$ for the total CO(1-0) emitting region, a 40\% correction factor accounting for emission resolved-out due to the high-resolution of our ALMA observations is included.  Column 7:  H$_2$ mass range assuming range assuming $4.6 < \alpha_{\mathrm{CO}} < 27$.}
\end{deluxetable*}

\begin{figure*}
\includegraphics[clip=true, trim=0.2cm 0cm 0cm 0cm, width=8.95cm]{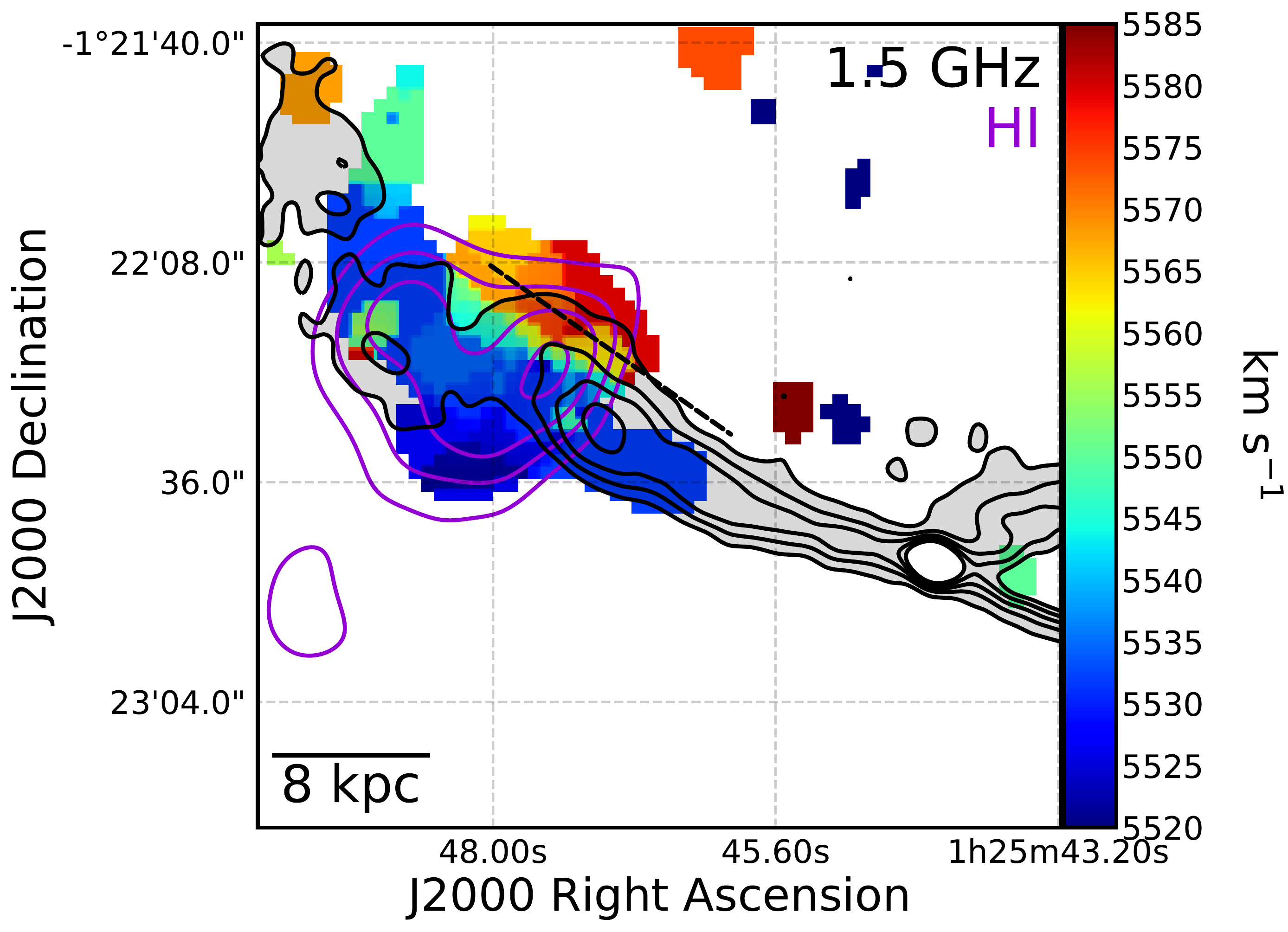}
\includegraphics[clip=true, trim=0cm 0cm 0.2cm 0cm, width=8.95cm]{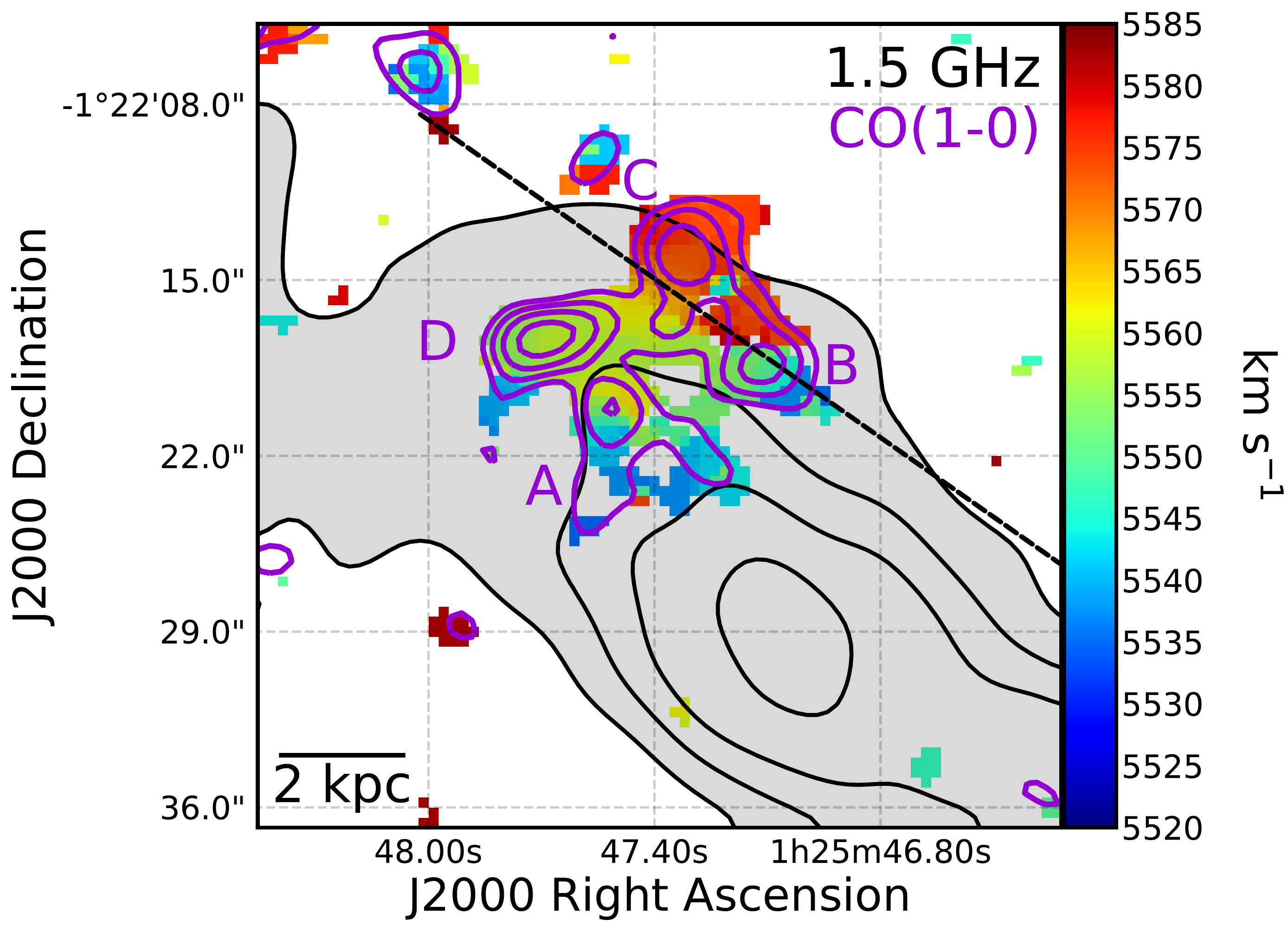}
\caption{Moment 1 (velocity) images in H{\tt I} and CO.  {\bf Left:} the H{\tt I} moment 1 image. The VLA 1.5~GHz continuum emission is represented by black contours at 1~mJy~beam$^{-1}$ intervals, starting from 1~mJy~beam$^{-1}$. Purple contours from the H{\tt I} moment 0 image are shown at levels of 50, 70, 90, 110 and 130~mJy~km~s$^{-1}$~beam$^{-1}$. {\bf Right:} the CO(1-0) moment 1 image. The VLA 1.5~GHz contours are the same as in the H{\tt I} image. Contours from the CO(1-0) moment 0 map are shown in purple at levels of 15, 20, 30 and 40~mJy~km~s$^{-1}$~beam$^{-1}$.  
The dashed black line in both images denotes the location of an abrupt velocity change possibly due to the interaction between the radio jet and the molecular gas.}
\label{fig:mom1}
\end{figure*}

\subsection{Molecular gas velocity structure}
In Figure~\ref{fig:mom1}, we present H{\tt I} and CO(1-0) moment 1 maps showing the 
gas velocity in comparison to the radio jet properties from the 1.5~GHz VLA data.  
The CO moment 1 image shows that the velocities of the CO clumps A, B, and D are all very similar and lie within a narrow velocity range of $\approx 5550-5560~$km~s$^{-1}$.  There is a small north-south velocity gradient across the southern clumps from 5540-5560~km~s$^{-1}$, then a more abrupt velocity jump to 5575~km~s$^{-1}$ in the vicinity of the northern clump C.  
Figure~\ref{fig:mom1} shows that this velocity jump is also seen in the  
H{\tt I} emission (in fact more clearly than in the CO emission), and the line along which it occurs is approximately aligned with the radio jet. This may thus represent dynamical evidence of the jet-cloud interaction. Much of the H{\tt I} emission lies to the South and East of the CO emission, and is at a lower velocity, $\approx$~5530~km~s$^{-1}$.

\begin{figure*}
\plotone{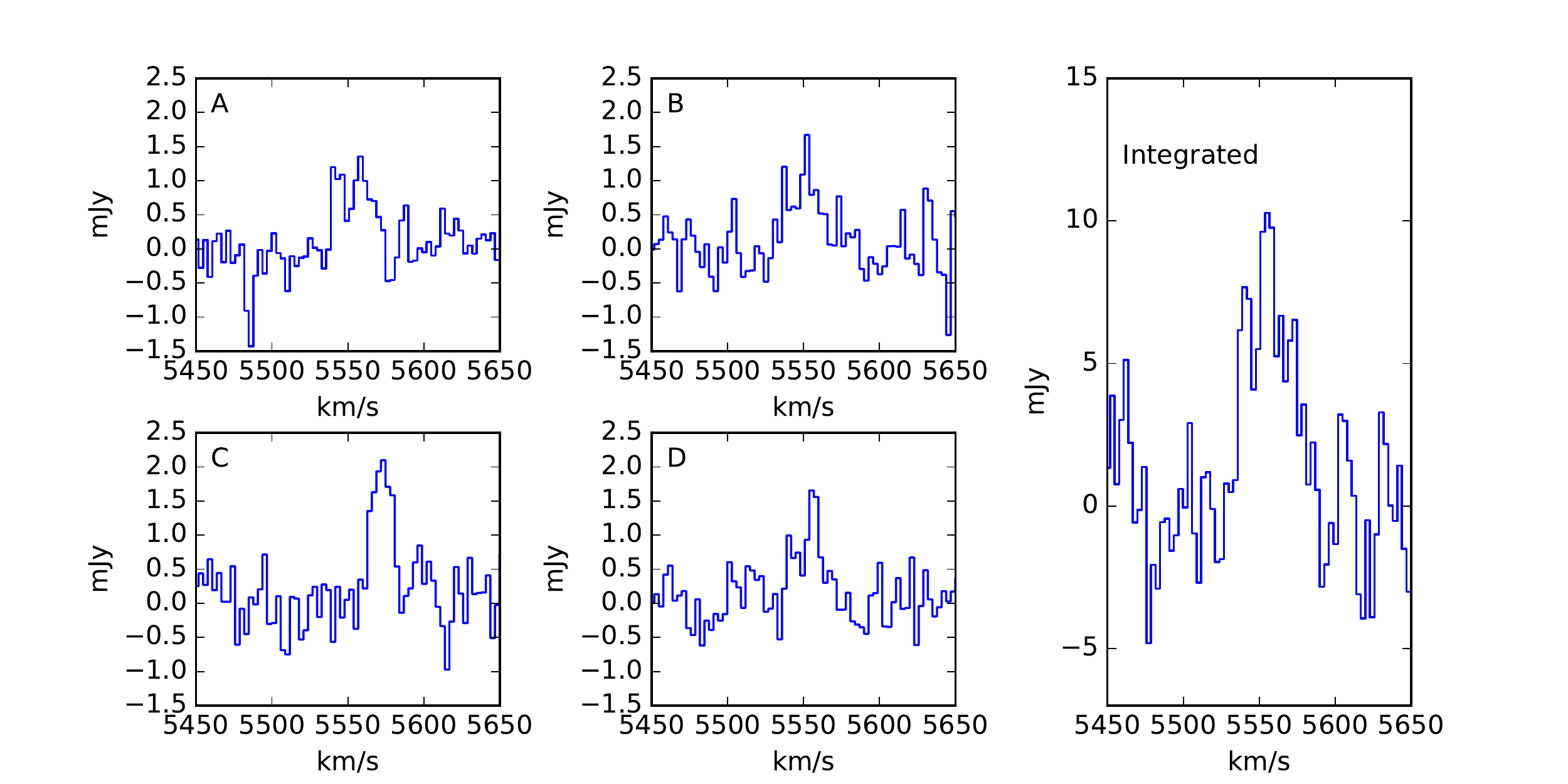}
\caption{{\bf Left:} spectra of CO clumps A-D (using the apertures in Table \ref{tab:gas_props}). {\bf Right:} Spectrum of the integrated CO emission from Minkowski's Object, using the integrated emission aperture in Table \ref{tab:gas_props}.}
\label{fig:spec}
\end{figure*}

\begin{figure*}
\includegraphics[clip=true, trim=0.2cm 0cm 0cm 0cm, width=8.95cm]{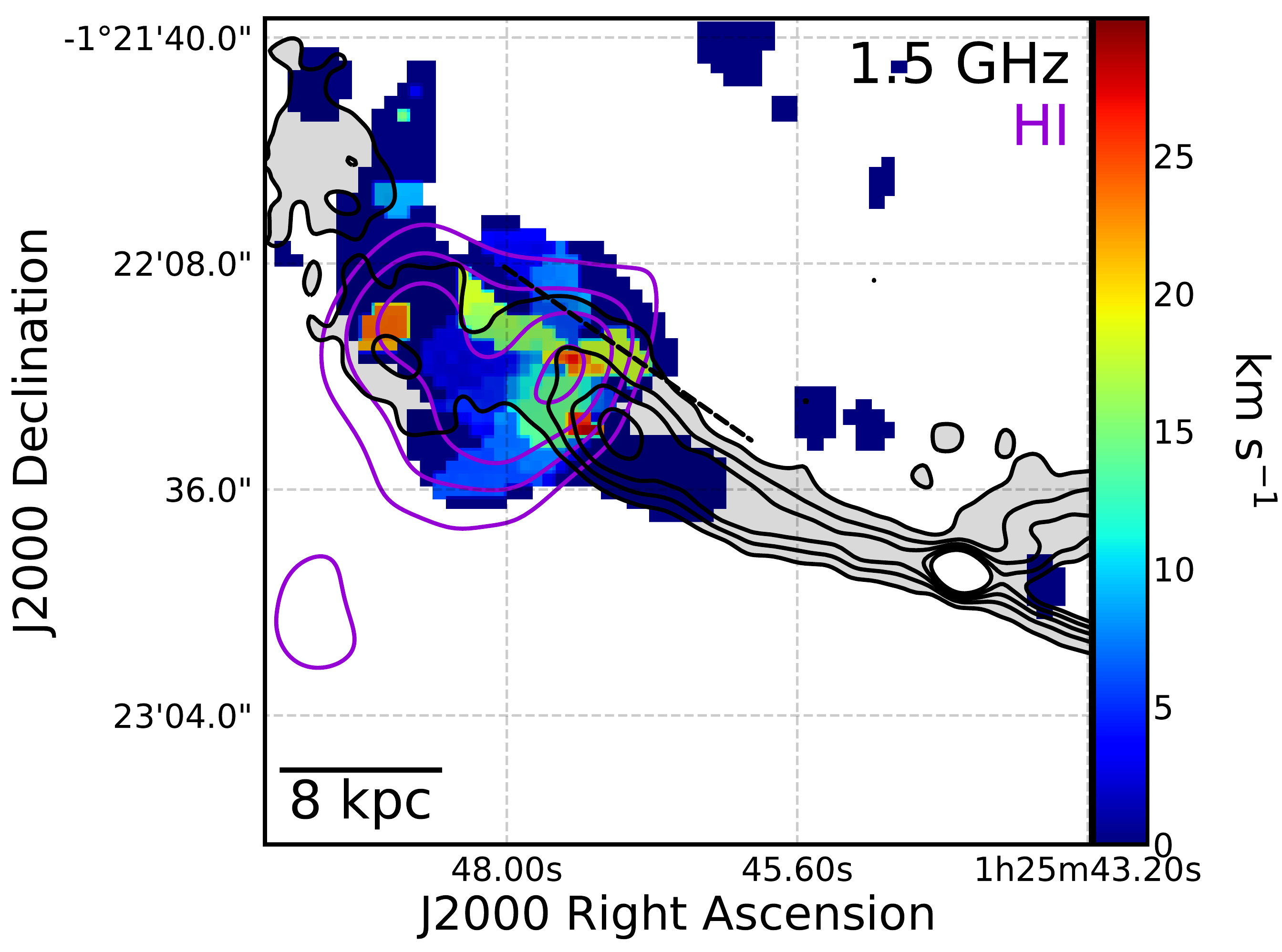}
\includegraphics[clip=true, trim=0cm 0cm 0.2cm 0cm, width=8.95cm]{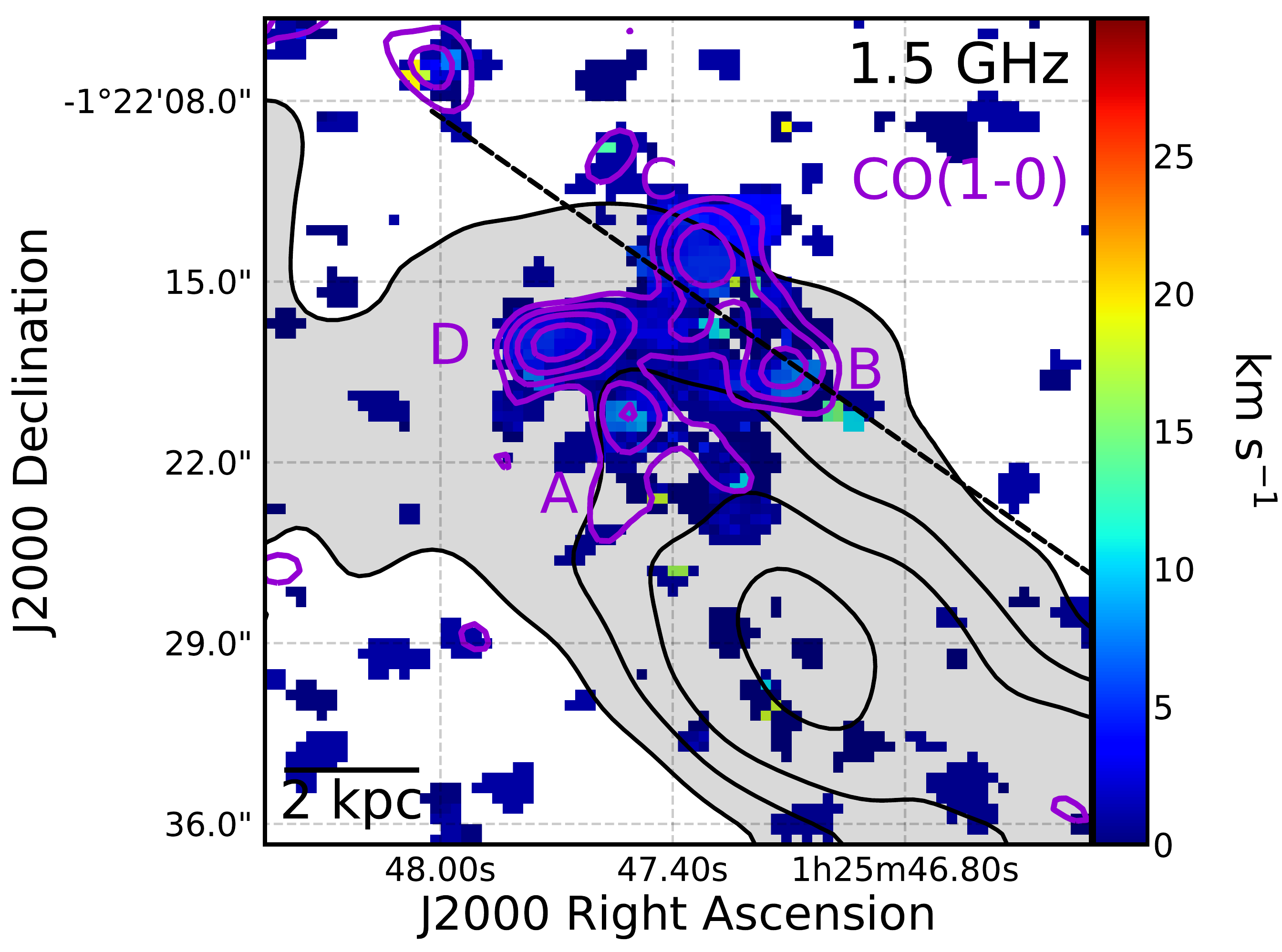}
\caption{
Moment 2 (velocity dispersion) images in H{\tt I} and CO.  {\bf Left:} the H{\tt I} moment 2 image. The black contours represent the VLA 1.5GHz emission and are at 1~mJy~beam$^{-1}$ intervals, starting from 1~mJy~beam$^{-1}$. Purple contours from the H{\tt I} moment 0 image are shown at levels of 50, 70, 90,110 and 130~mJy~km~s$^{-1}$~beam$^{-1}$. {\bf Right:} the CO(1-0) moment 2 image, with the extent of the VLA 1.5~GHz continuum emission indicated with the same contours as on the H{\tt I} image. Contours from the CO(1-0) moment 0 map are shown in purple at levels of 15, 20, 30 and 40~mJy~km~s$^{-1}$~beam$^{-1}$. 
The dashed black line in both images denotes the location of an abrupt velocity change possibly due to the interaction between the radio jet and the molecular gas.}
\label{fig:mom2}
\end{figure*}

Figure~\ref{fig:spec} shows the CO (1-0) spectrum of Minkoswski's Object and the four CO clumps (using the apertures defined in Table~\ref{tab:gas_props}), and Figure~\ref{fig:mom2} (right panel) shows the CO moment 2 map. 
Although the signal-to-noise is not very high, there is some evidence that the upstream knots (A and B) have higher velocity dispersions ($\approx 8 {\rm km\;s^{-1}}$) than the downstream knots C and D ($\approx 5~{\rm km\;s^{-1}}$), consistent with them having had more time to interact with the radio jet and develop more turbulent velocity fields as shocks propagate through the clouds, although 
in none of the clumps is the velocity dispersion very high. In contrast, as the left-hand panel of Figure~\ref{fig:mom2} shows, the H{\tt I} in the vicinity of the radio jet has a relatively high velocity dispersion, up to 25~${\rm km\;s^{-1}}$, consistent with being shocked by the passage of the radio jet.

\subsection{Molecular gas content}
We measure an integrated CO(1-0) flux density of $S_{\mathrm{CO}} = 289$~mJy~km~s$^{-1}$ within a 16$^{\prime \prime}$-diameter aperture centered at $\alpha_{\mathrm{J2000}} = 01{\rm h}25{\rm m}47.40{\rm s}$, $\delta_{\mathrm{J2000}} = -01^{\circ}22^{'}19.0{''}$ over the velocity range 5535-5586~km~s$^{-1}$ (barycentric, radio convention).  Because the large extent of the CO emission detected in Minkowski's Object, any emission present on even larger angular scales will be resolved out in our ALMA data.  Although the maximum recoverable scale (MRS) in our observations is $\approx 25^{\prime \prime}$, larger than the 16$^{\prime \prime}$ extent of the detected CO emission, flux recovery is known to fail on smaller scales than the MRS.  

To test whether we are likely to have underestimated the integrated CO flux density, we modeled the emission as a uniform disk of diameter 16$^{\prime \prime}$ and used the ALMA simulator in CASA to simulate the effects of spatial filtering by the ALMA interferometer.  The simulation showed that we may be underestimating the integrated CO flux density by as much as a factor of 1.4.  We therefore scale our measured 289~mJy~km~s$^{-1}$ flux by this amount and assume a 20\% uncertainty, resulting in an estimated flux density of $S^V_{\rm CO}=405\pm 81$~mJy~km~s$^{-1}$ for the CO(1-0) emission.  This can be compared to the limit of 0.032~mK~km~s$^{-1}$ in a 10~km~s$^{-1}$ channel (110~mJy~km~s$^{-1}$ in the 22$^{''}$ diameter IRAM beam) derived by Salom\'{e} et al.\ 2015 in CO(1-0).  The fact that the Salom\'{e} et al.\ (2015) limit from IRAM was considerably lower than our ALMA detection is most likely due to the smaller velocity width they used to obtain their limit ($10\;{\rm km~s^{-1}}$).

We convert the CO line flux to a luminosity $L_{\mathrm{CO}}$ via Equation~\ref{eq:Lco} below (Obreschkow et al. 2009):
\begin{equation}
\label{eq:Lco}
\begin{split}
\frac{L_{\mathrm{CO}}}{\rm K~km~s^{-1}~pc^2}=3.255 \times 10^7 \left(\frac{\nu_{\mathrm{o}}}{\rm GHz}\right)^{-2} \left(\frac{D_L}{\rm Mpc}\right)^2 \\
\times \, (1+z)^{-3} \frac{S_{\mathrm{CO}}}{\rm Jy~km~s^{-1}} = 6.5\times 10^6,
\end{split}
\end{equation}
where $D_L=82$~Mpc is the luminosity distance, $\nu_{\mathrm{o}}$ is the observed frequency of CO(1-0), $\nu_{\mathrm{o}} = \nu_{\mathrm{rest}}/(1+z)$~GHz, and $\nu_{\mathrm{rest}}$ is the rest frequency of the CO(1-0) transition at 115.27~GHz.

Inferring the molecular gas content of Minkowski's Object from our observations requires an estimate of the CO-to-H$_2$ mass conversion factor, $\alpha_{\rm CO}=M_{H_2}/L_{\rm CO}$, where $M_{H_2}$ is the molecular gas mass in solar masses and $L_{\rm CO}$ the CO luminosity in units of K~km~s$^{-1}$~pc$^2$. The usual assumption of Milky Way like conditions, in which $\alpha_{\rm CO} = 4.6 M_{\odot}{\rm (K~km~s^{-1}~pc^2)^{-1}}$ (Solomon et al. 1997), may not be appropriate for a low metallicity dwarf galaxy, so we also use the following prescription of Narayanan et al. (2012):
\begin{equation}
\label{eq:alpha}
\alpha_{\rm CO} =\frac{10.7 \langle W_{\rm CO} \rangle ^{-0.32}}{Z^{\prime \; 0.65}},
\end{equation}
where $Z^{\prime}$ is the metallicity of the gas relative to the sun and $<W_{\rm CO} >$ is the mean CO luminosity surface density in K~km~s$^{-1}$. We assume $Z^{\prime}=0.5$ based on the ionized gas metallicity (Croft et al. 2006) and the metallicity of the X-ray-emitting gas in the galaxy (Sakelliou et al. 2008), and an overall source area of 201~arcsec$^2$ corresponding to 29.2~kpc$^2$.  This results in a conversion factor of $\alpha_{\rm CO} = 27$~M$_{\odot}{\rm (K~km~s^{-1}~pc^2)^{-1}}$, which is significantly higher than the value usually assumed for the Milky Way. The estimated molecular gas mass corresponding to this conversion factor is $M_{\rm H_2}=1.8 \times 10^8$~M$_{\odot}$.  In contrast, if we instead assume the Milky Way conversion factor, the total molecular gas mass
estimate decreases to $M_{\rm H_2}=3.0 \times 10^7$~M$_{\odot}$. We assume these two values span the likely range for the molecular gas mass.  We find that clumps A-D each have masses consistent with the upper end of the mass range for giant molecular clouds observed in the Milky Way.  The individual CO clump molecular gas masses, as well as the total molecular gas mass, are summarized in Table~\ref{tab:gas_props}.  

We can compare our observed molecular gas fraction with the prediction of Fragile et al. (2004) for the molecular gas fraction ($f_{\rm mol}=M_{\rm H_2}/M_{\rm H{\tt I}}$) formed by the interaction of the radio jet and the H{\tt I} cloud. C06 give the atomic gas mass in H{\tt I} as $4.9\times 10^8 M_{\odot}$, so $f_{\rm mol}$ is in the range 6-36\%. Fragile et al. (2004) predict $f_{\rm mol} \sim 1$\% in the highest density clumps in their simulation, and much less ($\sim 10^{-3}$\%) overall, so it seems likely that there was a significant amount of pre-existing H$_2$ in Minkowski's Object before the interaction. Our estimate of $f_{\rm mol}$  is also consistent with the range seen in dwarf galaxies in the Virgo cluster (e.g., Grossi et al. 2016). 

\begin{figure*}
\includegraphics[clip=true, trim=0.2cm 0cm 0cm 0cm, width=8.95cm]{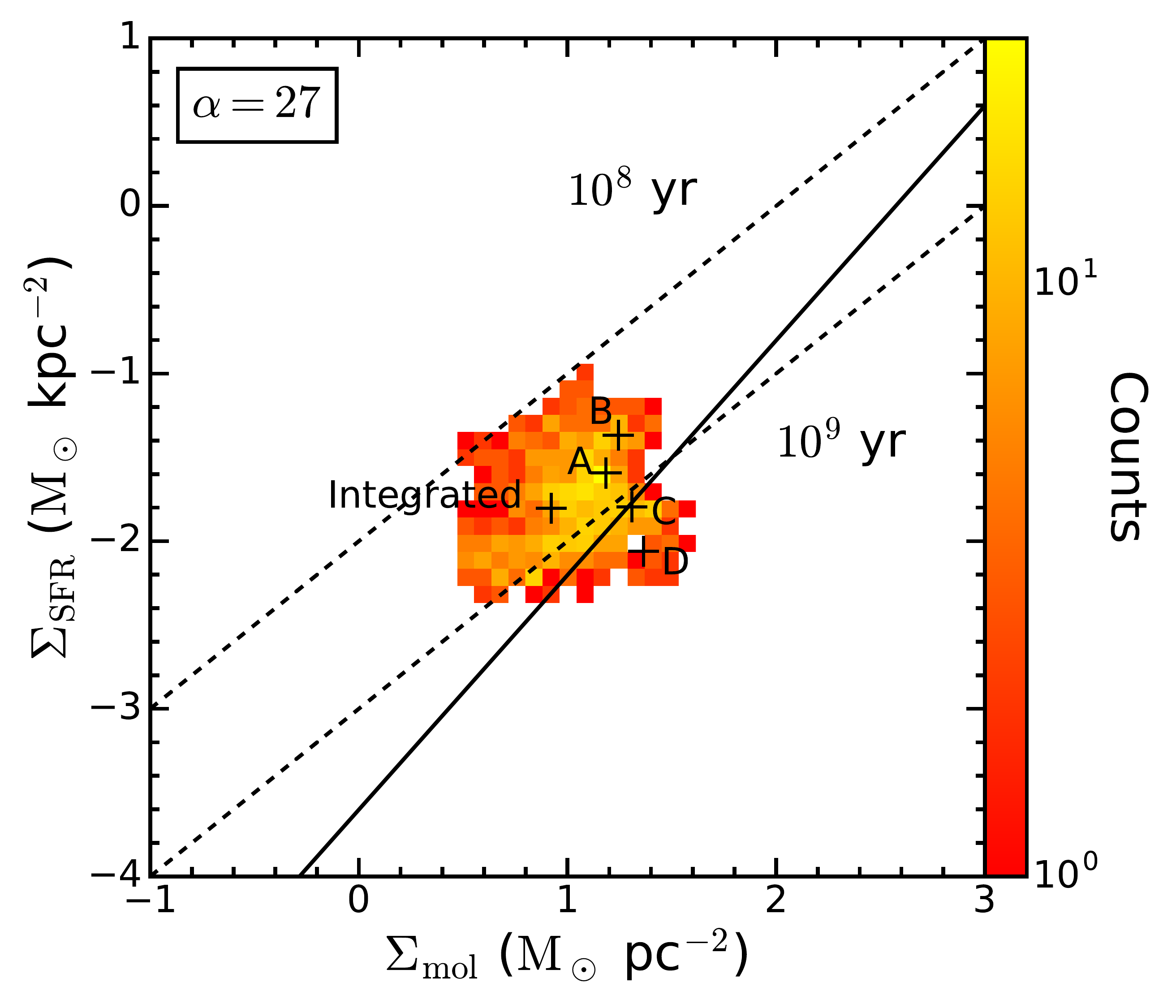}
\includegraphics[clip=true, trim=0cm 0cm 0.2cm 0cm, width=8.95cm]{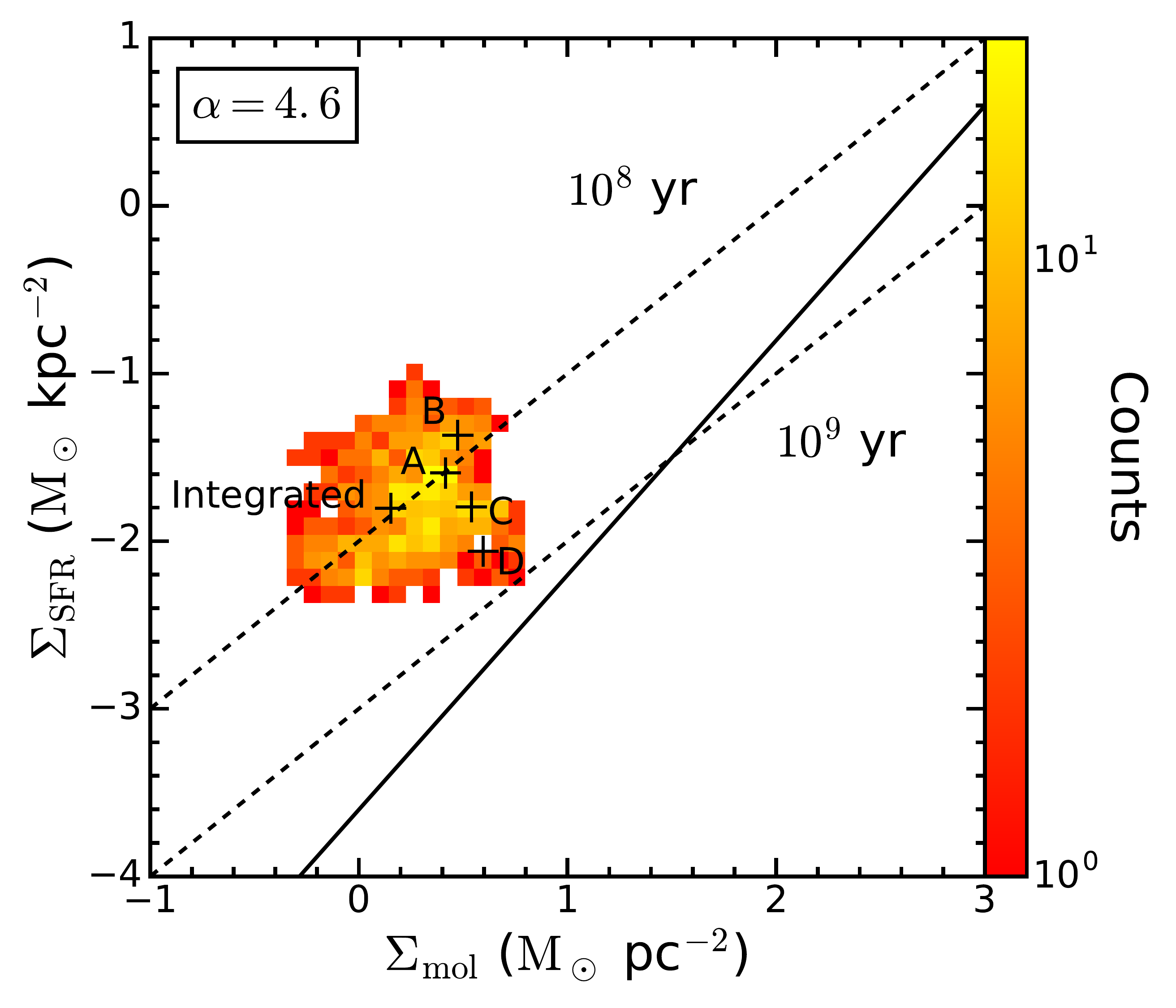}
\caption{{\bf Left:} Kennicutt-Schmidt relation plots of the star formation rate surface density ($\Sigma_{\mathrm{SFR}}$) as a function of molecular gas surface density ($\Sigma_{\rm H_2}$).  Here, we assume a value of the CO-to-H$_2$ conversion factor of $\alpha_{\mathrm{CO}} = 27$ appropriate for the low metallicity of the dwarf galaxy associated with Minkowski's Object.  The plots are pixel histograms, showing the range in values over the whole galaxy. The solid line indicates the relation for normal galaxies from Kennicutt (1998) and the upper and lower dotted lines correspond to gas depletion times of 10$^8$ and 10$^9$~yr, respectively.  {\bf Right:} Same as the left panel, except here we assume the standard Milky Way CO-to-H$_2$ conversion factor of $\alpha_{\mathrm{CO}} = 4.6$. The locations of clumps A-D and the integrated emission using the apertures in Table 1 are shown as crosses.}
\label{fig:ksplot}
\end{figure*}

\section{Discussion}

\subsection{Star formation rate and gas depletion time}
We constructed images of molecular gas surface density, $\Sigma_{\rm H_2}$, and star formation rate (SFR) surface density, $\Sigma_{\mathrm{SFR}}$, by scaling the CO(1-0) image and the H$\alpha$ image (from C06). Figure \ref{fig:ksplot} shows pixel histogram plots of $\Sigma_{\rm H_2}$ versus $\Sigma_{\mathrm{SFR}}$ for our two assumed values of $\alpha_{\rm CO}$. The values span the range of being above the Kennicutt-Schmidt relation by a factor $\sim 10$ in the main star-forming region, to being close to it in the vicinity of the downstream CO clump D.

\begin{figure*}
\plotone{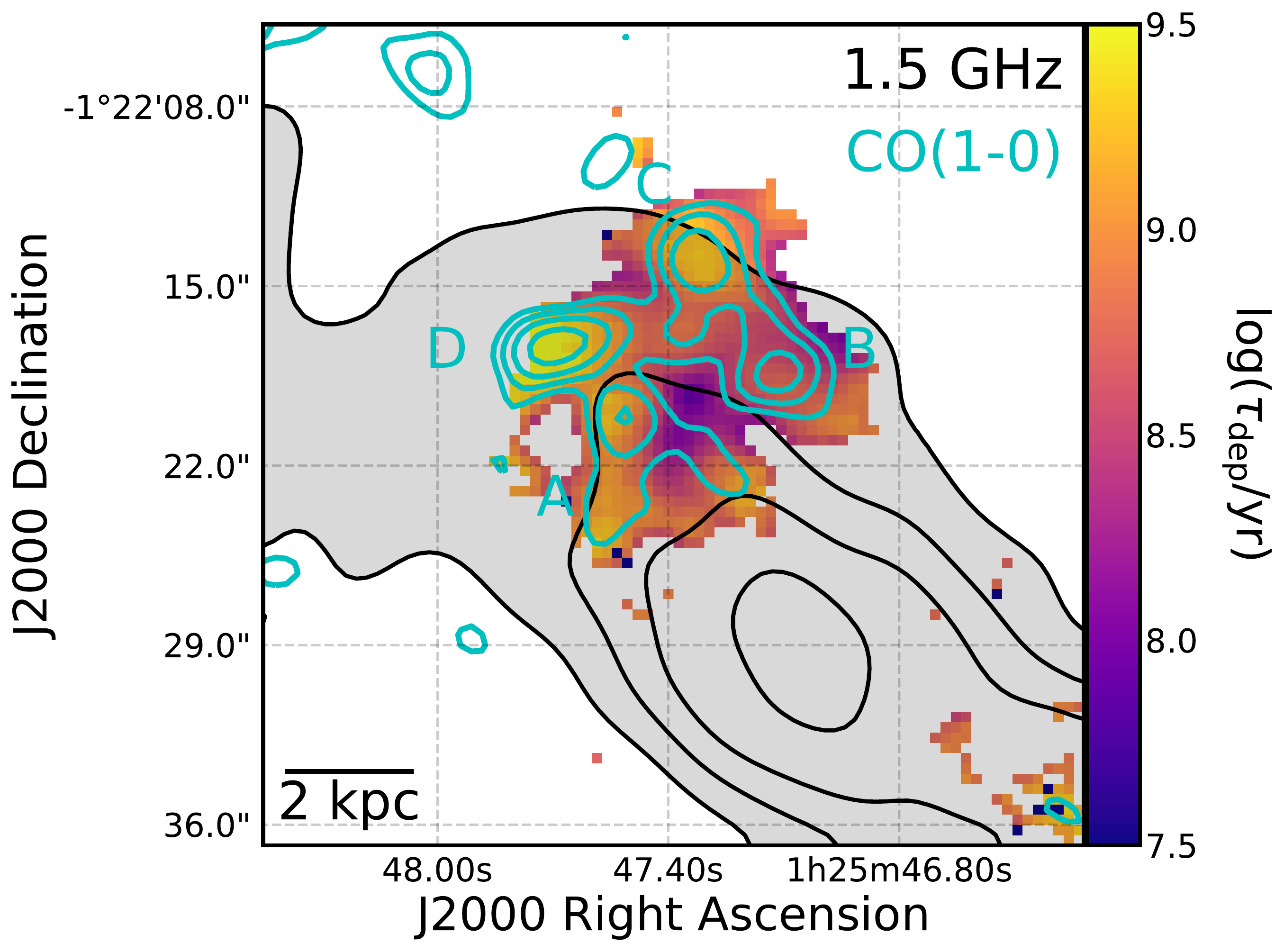}
\caption{Colorscale map of the gas depletion timescales for Minkowski's Object, assuming a CO-to-$H_{\mathrm{2}}$ conversion factor of $\alpha_{\mathrm{CO}} =27$ appropriate for the low-metallicity dwarf galaxy associated with Minkowski's Object.  To convert the depletion timescales corresponding to the canonical conversion factor for the Milky Way of $\alpha_{\mathrm{CO}} =4.6$, subtract 0.8 from the values shown in the colorbar.  Contours tracing the VLA 1.5~GHz data are overlaid in black and shown at 1~mJy~beam$^{-1}$ intervals, starting from 1~mJy~beam$^{-1}$.  Contours from the CO(1-0) moment 0 map are shown in cyan at 15, 20, 30 and 40~mJy~km~s$^{-1}$~beam$^{-1}$, and clumps A-D are labeled.
\label{fig:tdimage}}
\end{figure*}

For Minkowski's Object as a whole, we assume a SFR of 0.47~M$_{\odot}$~yr$^{-1}$ (Salom\'{e} et al. 2015), this leads to a short average gas depletion timescale $\sim 0.6-3.8\times 10^8$~yr. In Figure \ref{fig:tdimage}, we plot an image of the gas depletion time (calculated simply as the ratio of gas surface density divided by star formation rate surface density), showing a rise in the downstream direction towards the more intact CO clumps (C and D), where presumably star formation is yet to begin in earnest.

We can compare this timescale to estimates of the age of the radio AGN.  Sakelliou et al. (2008) estimate an age based on the spectral index at the extremes of the radio plumes of both radio sources in the cluster (3C40A, associated with the radio jet in NGC541, and 3C40B associated with NGC547) of $\approx 1.4 \times 10^8$~yr. Independently, Bogd\'{a}n et al. (2011) estimate an age for the radio AGN associated with 3C40B based on the time to evacuate the X-ray cavity around its southern lobe of $8\times 10^7$~yr.  Thus, estimates for the age of the radio source are consistent the gas depletion time and greater than the age of the starburst ($\sim 10^7$~yr; C06).

Note that the above estimates of star formation surface density assume a normal initial mass function (IMF) for star formation in Minkowski's Object. The recent explosion of a Type II supernova, SN~2010ib (Cenko et al. 2010; Gal-Yam et al. 2010), however, hints at a top heavy IMF. The star formation rate in Minkowski's Object is only $\approx 0.5$~M$_{\odot}$~yr$^{-1}$, and automated supernova searches have only been ongoing for $\sim$20 years, so observing a supernova is very unlikely unless the IMF is skewed towards high mass stars. Rauch et al. (2013) also argue for a top-heavy IMF from jet-induced star formation in the companion of a high redshift quasar. If the IMF were to be top heavy, then the actual star formation rate would be lower than estimated above, and thus Minkowski's Object would lie closer to the Kennicutt-Schmidt relation.

\subsection{Physical conditions in the ISM}

We use the VLA 1.5~GHz data to estimate the pressure $p_{\rm me}$  corresponding to the minimum energy density $u_{\rm me}$ required to produce the observed synchrotron emission in the interaction region, using the standard formulation (e.g. Miley 1980):
\begin{equation}
p_{\rm me}= \frac{u_{\rm me}}{3} = \frac{7}{9} \frac{B_{\rm me}^2}{2 \mu_0},
\end{equation}
where
\begin{eqnarray}
B_{\rm me}&=&5.69\times 10^{-9} \times  \nonumber \\
&&\left[\frac{(1+\kappa)}{\eta}(1+z)^{(3-\alpha)}\frac{1}{\theta_x \theta_y\; s \; {\rm sin}^{1.5}\phi}\frac{S_0}{\nu_0^{\alpha}}\frac{\nu_2^{\alpha+0.5}-\nu_1^{\alpha+0.5}}{\alpha+0.5}\right]^{\frac{2}{7}}
\end{eqnarray}
$\mu_0$ is the permeability of free space, $\kappa$ is the ratio of energy in heavy particles to that in electrons (assumed to be 100), $\eta$ is the filling factor of the emission region (assumed to be 1), $\theta_x=4.2$~arcsec and $\theta_y=3.8$~arcsec correspond to the size of the beam, $s=4.0$~kpc is our estimate of the path length through the source, $\phi$ is the angle between the magnetic field and the line of sight (assumed to be 45$^{\circ}$), $S_0=0.0018$~Jy/beam is the mean surface brightness of the emission, $\alpha=-0.8$, $\nu_0=1.5$~GHz is the frequency of observation, and $\nu_1=120$~GHz and $\nu_2=0.3$~GHz are the assumed upper and lower cutoffs in the synchrotron spectrum. Using these equations we find a minimum pressure of $1.3 \times 10^{-12}$~Pa. This is comparable to the estimate from the X-ray gas in Abell 194 by Bogd\'{a}n et al. (2011), $1.7\times 10^{-12}$~Pa, and suggests that the interaction region is not highly overpressured compared to the surrounding gas, consistent with the idea that the radio jet interaction is only inducing weak shocks in the ISM. We can compare the pressures in the interaction region to those in the jet just before it impacts the galaxy.  Assuming equipartition, the ratio is given by the ratio of the surface brightness of the synchrotron emission to the 4/7 power. The peak surface brightness in the final jet knot (centered on 
$\alpha_{\mathrm{J2000}} = 01{\rm h}25{\rm m}47.0{\rm s}$, $\delta_{\mathrm{J2000}} = -01^{\circ}22^{'}29{''}$) is $\approx 4.5$~mJy, so the pressure change is a factor of $\sim (4.5/1.8)^{4/7}=1.7$. These weak shocks could still have a significant effect on the ISM, and may in fact be more efficient at coupling to the ISM than stronger shocks (e.g. Appleton et al. 2013). 

There is a small amount of X-ray emission from Minkowski's Object that has been ascribed to high-mass X-ray binaries (Bogd\'{an} et al. 2011), but this requires a higher star formation rate than observed.
We speculate that some of the X-ray emission could be from shocks, or else be further evidence for a top-heavy IMF.

\subsection{Background starburst galaxy}
The ALMA continuum image reveals a compact source at $\alpha_{\mathrm{J2000}} = 01{\rm h}25{\rm m}46.89{\rm s}$, $\delta_{\mathrm{J2000}} = -01^{\circ}22^{'}10.2{''}$ with a peak flux density of $S_{\rm 106\,GHz}=118 \pm 8 \mu{\rm Jy}$ that is not present in the VLA 1.5~GHz image. This is most likely to be a background high-redshift galaxy.  Inspection of the data cubes reveals a single strong emission line at 110.5~GHz with a double-horned profile.  Given that the source is absent from the HST data, despite being bright in both archival images from the {\it Spitzer Space Telescope} and at 3~mm in our ALMA observations, we can probably eliminate CO(1-0) line emission at $z=0.04$. The most likely line identifications are thus CO(2-1), corresponding to $z=1.09$, or CO(3-2) at $z=2.13$.


\section{Summary and Future Work}

We have made the first detection of molecular gas in Minkowski's Object using ALMA. Our estimate of  $M_{\rm H_2} =3.0 \times 10^7~M_{\odot}$ for the Milky Way value of $\alpha_{\rm CO}=4.6~M_{\odot}{\rm (K~km~s^{-1}~pc^2)^{-1}}$ is higher than the limit of $1 \times 10^7~M_{\odot}$ found by Salom\'{e} et al. (2015) with the IRAM 30m for the same value of $\alpha_{\rm CO}$, most likely because we integrated over a larger velocity range (31 km$\;$s$^{-1}$ compared to 10 km$\;$s$^{-1}$).

The results of our study are fully consistent with the presence of an interaction between the radio jet and molecular gas in Minkowski's Object that is triggering and/or boosting star formation.
First, star formation is enhanced on average compared to the Kennicutt-Schmidt relation (Figure \ref{fig:ksplot}). Second, the star formation rate surface density is higher and the gas depletion time shorter in the CO clumps in the upstream direction of the radio jet (clumps A and B) where the jet has influenced the molecular gas for a longer period of time, than in the downstream clumps C and D. This is also consistent with the evidence we find for higher velocity dispersions in the molecular gas in clumps A and B. Third, we also find marginal evidence for a flattening of the radio spectral index in the vicinity of the detected CO clumps that is suggestive of cosmic ray electron re-acceleration in regions where the radio jet has driven shocks in the molecular gas.  In addition, the H{\tt I} images presented in this paper show a high velocity dispersion in the region of the H{\tt I} cloud in the vicinity of the radio jet, consistent with jet-induced shocks, and both the H{\tt I} and CO emission show signs of a small ($\approx$15~km~s$^{-1}$), but abrupt change in velocity along the edge of the jet near CO clump C (see Figure~\ref{fig:mom1}), that could be interpreted as further dynamical evidence for the presence of jet-disrupted atomic and molecular gas in Minkowski's Object.

The timescales we measure are also consistent with jet-triggered star formation.
The upstream regions have short gas depletion timescales $\sim 10^8$~yr, longer than the age of the starburst as estimated by C06 ($\sim 10^7$~yr), and longer still than the cloud compression timescale from the simulations of Fragile et al. (2004), $\sim 10^6$~yr.  Another important timescale is the time taken for Minkowski's Object to cross the path of the jet. The radial velocity difference between Minkowski's Object and NGC541 is about 250~km~s$^{-1}$, at which speed Minkowski's Object would take $\approx 2\times 10^7$~yr to traverse the path of the jet.  All of these timescales are shorter than the likely lifetime of the radio source of $\sim 10^8$~yr.  These timescales are thus consistent with the scenario in which Minkowski's Object (whether it formed from a gas cloud in situ, or originated from a dwarf galaxy infalling into the cluster) experienced a burst of star formation as it crossed the path of the radio jet in an event that started $\sim 10^7$~yr ago, and has sufficient molecular gas to continue forming stars for the duration of the interaction. 

The ultimate fate of Minkowski's Object is an interesting question given our current understanding of AGN feedback.  Although we find compelling evidence that Minkowski's Object is indeed an example of positive radio jet feedback, it is unlikely that the stars formed during the interaction will accrete onto NGC541, the massive early-type galaxy hosting the radio AGN responsible for the feedback.  Instead, we think it more likely that Minkowski's Object will remain bound and become a dwarf galaxy in the Abell 194 cluster, or it will be dispersed due to supernova feedback and tidal effects, with its stars joining the intracluster stellar population (e.g.\ Mihos et al. 2005).  So the net effect of interactions like this may result in negative feedback on the host of the AGN itself, by preventing further accretion of gas that could then cool and lead to increased star formation in the host.  Although cases like this are rare in the local Universe, young clusters like Abell 194 were much more common at $z\sim 1$, and therefore, when considered on cosmological timescales, both positive and negative radio AGN feedback are likely to be important feedback mechanisms in regulating the growth and evolution of galaxies.   

This paper is part of a new series of papers on Minkowski's Object that will examine in detail the interaction between the radio jet and gas using both new numerical simulations (Fragile et al.\ 2017) and new observations with the Very Large Array and {\em Hubble Space Telescope}. With these new simulations and data we will further test the hypothesis of jet-induced star formation in Minkowski's Object, and study the detailed physics of the interaction.

\section*{Acknowledgments}
This paper makes use of the following ALMA data: ADS/JAO.ALMA\#2015.1.00570.S. ALMA is a partnership of ESO (representing its member states), NSF (USA) and NINS (Japan), together with NRC (Canada), NSC and ASIAA (Taiwan) and KASI (Republic of Korea) in cooperation with the Republic of Chile. The joint ALMA Observatory is operated by ESO, AUI/NRAO and NAOJ. The National Radio Astronomy Observatory is a facility of the National Science Foundation operated under cooperative agreement by Associated Universities Inc. This research made use of Astropy, a community-developed core Python package for Astronomy (Astropy Collaboration, 2013). This research made use of Montage. It is funded by the National Science Foundation under Grant Number ACI-1440620, and was previously funded by the National Aeronautics and Space Administration's Earth Science Technology Office, Computation Technologies Project, under Cooperative Agreement Number NCC5-626 between NASA and the California Institute of Technology. This research made use of APLpy, an open-source plotting package for Python (Robitaille and Bressert, 2012).




\end{document}